\documentclass[12pt]{iopart}
\usepackage{iopams}
\usepackage{graphicx}
\usepackage{subfigure}
\usepackage{cancel}
\usepackage{ulem}

\begin{document}

\title[Quasi-stationary simulations of DP class in $d=3$]{Quasi-stationary simulations of the directed percolation universality class in $d=3$ dimensions}

\author{Renan S Sander$^1$, Marcelo M de Oliveira$^{1,2}$ and Silvio C Ferreira$^1$}
\address{$^{1}$Departamento de F\'{\i}sica, Universidade Federal
de Vi\c{c}osa, 36571-000, Vi\c{c}osa - MG, Brazil\\$^{2}$Campus Alto
Paraopeba, Universidade Federal de S\~ao Jo\~ao Del Rei, 36420-000,
Ouro Branco - MG, Brazil. }
\ead{silviojr@ufv.br}

\begin{abstract}
We present quasi-stationary simulations of three-dimensional models
with a single absorbing configuration, viz. the contact process
(CP), the susceptible-infected-susceptible (SIS) and the contact
replication process (CRP). The moment ratios of the order parameters
for DP class in three dimensions were set up using the well
established SIS and CP models. We also show that the mean-field
exponents in $d=3$ reported previously for CRP [Ferreira SC 2005 \PR
E \textbf{71} 017104] is a transient observed in the spreading
analysis.

\end{abstract}

\pacs{05.70.Ln, 05.65.+b, 02.50.Ey}
\submitto{\it J. Stat. Mech.: Theor. Exper.}
\maketitle

\section{Introduction}

Phase transitions to a single absorbing configuration, a state in
which the system can not scape from, are nowadays a topic in the
frontier of Nonequilibrium Statistical Physics
 \cite{Marro,Odor}. Concomitantly with the increasing interest on absorbing/active
 phase transitions in complex topologies \cite{BookRedes,Pastor,Ferreira2007,Colizza,voronoi},
 there are still a lot of open problems being intensively investigated on regular lattices
 such as the effects of quenched disorder \cite{vojta,dahmen,Oliveira2008}, diffusion
 \cite{difusao}, as well the modeling of predator-prey systems \cite{Arashiro}, and clonal replication \cite{Ferreira2004,Ferreira2005}.

Under the renormalization group point of view, it is expected
\cite{Marro,jans81,gras82} that the absorbing phase transitions in
models with a positive one-component order parameter, short-range
interactions and without additional symmetries or quenched disorder
belong generally to the universality class of directed percolation
(DP). This conjecture is known as Janssen-Grassberger criterion
\cite{Marro}. It is worthwhile to mention, the interest on this kind
of phase transitions was raised by the recent experimental
observation of the DP class in absorbing-state phase transitions
\cite{take07,pine}. On the other hand, while DP is considered the
most robust universality class of the absorbing-state phase
transitions, the precise numerical determination of the critical
exponents of a specific model can be masked by  factors like
diffusion\cite{difusao} and weak quenched disorder \cite{Oliveira2008}.

The contact process (CP), the standard example of the DP
universality class, is a toy model of epidemics \cite{Harris}.\footnote{A CP variation called susceptible-infected-susceptible
(SIS) model is more widely applied in epidemiological studies
\cite{BookRedes}.}. More recently, a novel variation of the CP was
introduced for the modeling of clonal (copies of
themselves) replication, the contact replication process (CRP)
\cite{Ferreira2004,Ferreira2005}. Since neither additional
symmetries nor long-range interactions were included, the CRP
fulfils the requirements of the Janssen-Gassberger criterion.
However, the first dynamic spreading analysis of CRP in $d=1-3$
dimensions, reported in \cite{Ferreira2004,Ferreira2005}
intriguingly classified the model in the DP universality class in
one and two, but not in three dimensions. Surprisingly, in $d=3$ the
reported spreading exponents were those predicted by the mean-field
approach \cite{Ferreira2005}.

In the present work we applied spreading analysis and the method of
quasi-stationary simulations \cite{Vidigal,Oliveira2005} in
three-dimensional models that fulfill the Janssen-Grassberger
criterion. Particularly, we turned back to the CRP model and showed
that the mean-field behaviour observed previously  in $d=3$
\cite{Ferreira2005} is a transient associated to the closeness
between critical creation and annihilation events. Additionally, we
have analysed several moment ratios in $d=3$ and have determined the
universal values for DP class based on the results obtained from CP
and SIS models.  The paper is outlined as follows. In section
\ref{model}, the models and simulation procedures are described.
Simulation results are presented and discussed in section
\ref{result}. Conclusions are
drawn in section \ref{conclusion}.

\section{\label{model}Models and methods}
\subsection{Models}
The \textit{contact process} (CP) is defined in a hypercubic lattice
$\mathcal{Z}^d$ in which the sites represent individuals in two
states: healthy ($\mathcal{Z}_i = 0$) or infected ($\mathcal{Z}_i =
1$). Infected sites become healthy (empty) at unitary rate $1$ while
healthy sites are infected at a rate $n_i\lambda/q$, where $q$ is
the lattice coordination number and $n_i$ the number of infected
sites surrounding (first-neighbours in the distance) the healthy site
$i$. These rules implies, for values of infection rate above a
certain $\lambda_c$, an infection flowing equally distributed among
all nearest neighbours (NN) of infected sites.

The \textit{susceptible-infected-susceptible} model is a variation
of the CP dynamics in which any empty site with one or more infected
nearest-neighbours becomes infected at rate $\lambda$. Again, a
unitary cure rate is assumed. In the literature, the SIS model is
also known as the A model \cite{jaff}.

\textit{Contact replication process} rules are are very similar to
those of CP. Instead of individuals, the sites represent places
where cells lie. Analogously to the spontaneous cure in CP, cells
die at unitary rate. However, a cell replicates at a rate $\lambda$
and the offspring occupies one of its empty NN chosen at random. So,
an empty site $i$ is occupied at rate $\lambda \sum_j
\mathcal{Z}_j/n_j$. The sum is done over all neighbours of the site
$i$ and $n_j$ is defined as before. Notice that the creation process
is facilitated in the CRP in relation to CP since the occupation
flows uniformly among only empty neighbours implying  lower critical
rates. Indeed,  estimates reported for CRP were
$\lambda_c=2.02634(4), 1.08320(7)$, and $1.0000(1)$ for $d=1,2,$ and
3 \cite{Ferreira2004, Ferreira2005}, in comparison with those for
the CP $\lambda_c=3.29785(2),1.64877(3),$ and 1.31686(1)
\cite{rewetting}, respectively. Notice that the three models share
the same symmetries and, consequently, they are expected to belong
to the same universality class.

For all these models, Monte Carlo simulations were performed using
the usual procedure \cite{Marro}: An event, creation or
annihilation, is selected with probabilities $p=\lambda/(1+\lambda)$
and $1-p$, respectively,  and an occupied site $i$ is chosen at
random. In the annihilation process, the occupied site become empty
in all models while the creation depends on model. In CP, one
nearest-neighbour (NN) of $i$ is chosen at random and infected if
empty, otherwise nothing occurs. In SIS, all empty sites
neighbouring the site $i$ are occupied. Finally, in CRP one of empty
neighbours, if there are anyone, is chosen at random and occupied.
In all cases, the time is incremented by $\Delta t = 1/N$, where $N$
is total number of infected sites.

\subsection{\label{qs}Quasi-stationary simulations}

Stationary analysis of systems with transitions to absorbing
configurations in the proximity of the critical point are ruled by
strong finite size effects. Indeed, the unique actual stationary
state of finite systems is the absorbing one. A common alternative
to avoid this difficulty is to restrict the averages to the survival
samples and apply a finite size analysis. However, such procedure is
not free of ambiguities or misinterpretations \cite{Oliveira2005}.
An alternative approach is the quasi-stationary QS simulation method
\cite{Vidigal,Oliveira2005}. This method consists of storing a list
with $M$ configurations visited in the history of the system and
periodically replacing one of them by the current state. Whenever
the system try to visit the absorbing state, the configuration is
replaced by an active one selected at random from the list
containing the sample of configurations and the simulation continues
as usually.

The QS simulations were performed as follows. Firstly, the list of
configurations is incremented whenever the time increases by a unity
up to a list with $M$ configurations is achieved. Secondly, a
configuration of the list randomly chosen is replaced by the current
one with a given probability $p_{rep}$. We used a large value of
$p_{rep}=0.05$ for a initial relaxation period, more precisely for
$t<5\times 10^7$ aiming to speed up the erasing of the memory of the
initial conditions. In turn, $p_{rep}=2\times10^ {-5}$ was adopted
for the remaining of the simulation. Runs with $t_m=2\times 10^ 8$
steps and averages after a relaxation time $t_r=1\times 10^ 8$ were
used. Finally, the averages and uncertainties were obtained with at
least $15$ (to the largest system) independent runs with a full
lattice as initial condition. Periodic boundary condition were
always used.

\subsection{\label{spreading}Spreading analysis}

The critical point $\lambda_c$ can be efficiently determined by the
spreading analysis, which consists of evolving the system from a
perturbation to the absorbing state (a single occupied site at the
origin of the lattice) and computing the survival probability $P$
and mean number of occupied sites $N$ as time functions. In this
analysis, the averages are done over all samples, surviving or not.
Asymptotic power law dependencies,
\begin{equation}
P(t)\sim t^{-\delta} ~ \mbox{and}~N(t)\sim t^\eta
\end{equation}
are expected at criticality. Since $N\sim t^d$ at upper-critical and exponentially
decays in the sub-critical regimes, deviations from power laws are expected around
the critical point and a null curvature criterion of the double-logarithm plots of $N$
against $t$ can be used to determine the critical point \cite{Ferreira2004}.
Analogous analysis can be done with the survival probability.

The spreading is defined by
\begin{equation}
R^2(t) = \frac{1}{N(t)}\left\langle \sum_{j\in\mathcal{Z}^d} r_j^ 2 \mathcal{Z}_j\right\rangle,
\end{equation}
in which $r_j$ is the distance from the original seed where the perturbation was introduced.
At criticality, it is expected an asymptotic power law
\begin{equation}
R^2(t)\sim t^z.
\end{equation}
The spreading exponents obey a hyperscaling relation $4\delta+2\eta
= d z$ \cite{Marro,Odor}.

\begin{figure}[hbt]
\begin{center}
\includegraphics[width=9.0cm,height=!,clip=true]{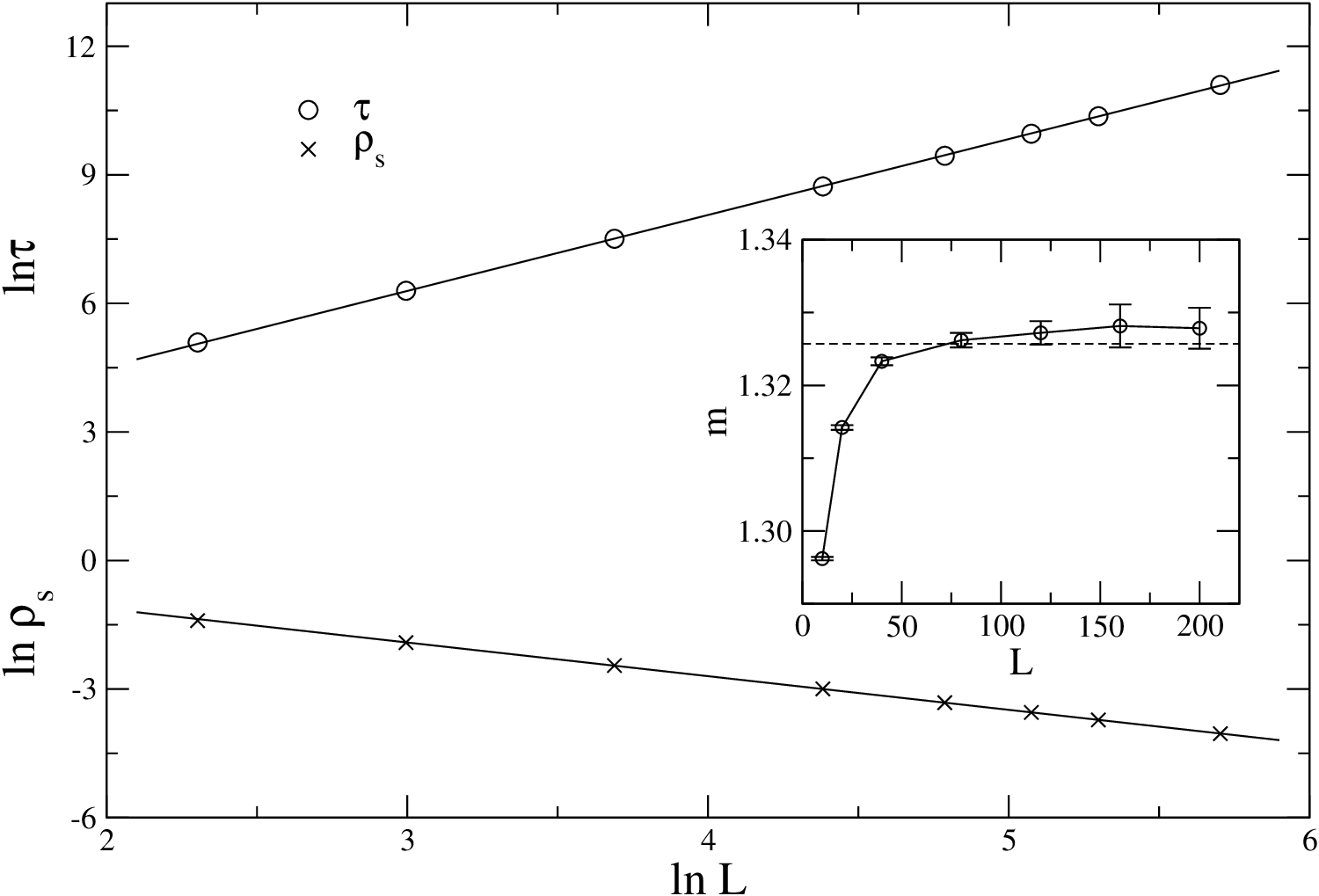}
\caption{\label{fig1} Critical QS density $\rho_s$ and lifetime
$\tau$ versus system size $L$ for the two-dimensional CRP model.
Inset: critical moment ratio
$m=\langle\rho^2\rangle/\langle\rho\rangle^2$ for varying system
sizes. Dashed line is the moment ratio for the contact process taken
from \cite{jaff}.}
\end{center}
\end{figure}

In order to illustrate the method, we present results from
quasi-stationary simulations of the two-dimensional CRP model\footnote{in ref. 
\cite{Ferreira2005} only results from spreading
simulations were reported.}. The QS density vanishes as usual at
$\lambda=\lambda_c=1.08322(2)$ when $L\rightarrow\infty$ (the
critical rate were determined using the spreading analysis). Figure
\ref{fig1} shows the critical densities and lifetimes of the QS
state {versus} system size. Lifetime was calculated as
\cite{Vidigal}
\begin{equation}
\tau=1/p_1,
\end{equation}
where $p_1$ is the probability of attempting the absorbing
configuration in QS simulations. The power laws
\begin{equation}
\rho_s\sim L^{-\beta/\nu_\perp}
\end{equation}
and
\begin{equation}
\tau\sim L^{\nu_\parallel/\nu_\perp}
\end{equation}
were verified and the slopes for $L>20$ provide $\beta/\nu_\perp =
0.800(7)$ and $\nu_\parallel/\nu_\perp = 1.764(14)$. As expected,
these exponents are in very good accordance with those reported for
the DP universality class \cite{rewetting} (table \ref{exponents}).
Also, the critical variance of the order parameter, defined by $\chi
= L^d(\langle\rho^ 2\rangle - \langle\rho\rangle^ 2)$ \cite{Marro},
scales as
\begin{equation}
\chi \sim  L^{\gamma/\nu_\perp},
\end{equation}
with an exponent $\gamma/\nu_\perp = 0.412(8)$, again agreing with
the DP value.

The ratio between moments of the order parameter is widespread as an
useful tool for the determination of the critical point of
equilibrium systems \cite{Landau}. This concept has been extended to
non-equilibrium systems, particularly to models belonging to the DP
universality class for which the moment and cumulant ratios have
proven to be universal quantities \cite{jaff}.  The inset of figure
\ref{fig1}(b) shows the ratio $m = \langle \rho ^2\rangle / \langle
\rho \rangle^2$ for the two-dimensional CRP model at criticality for
varying system sizes. It was shown that this ratio assumes the
critical value $m_c=1.3257(5)$ for CP in two dimensions \cite{jaff}.
Using data for $L\ge 80$ we found $m_c = 1.3274(9)$ for
two-dimensional CRP, again in accordance with DP universality class.
Additionally, we also verified the agreement of higher moment ratios
with the predictions for DP classes for $d=1$ and 2.

\begin{table}[hbt]
\caption{\label{exponents}Critical exponents for DP class calculated
from spreading exponents given in \cite{rewetting}. Scaling
relations $\beta = \delta \nu_\parallel$,
$z=2\nu_\perp/\nu_\parallel$, and $\gamma=d\nu_\perp-2\beta$
\cite{Marro} were used whenever necessary.}
\begin{indented}
\item[]\begin{tabular}{cccccc}
\br \multicolumn{3}{c}{Quasistationary} &
\multicolumn{3}{c}{Spreading} \\ \br
Exponent& $d=2$ &  $d=3$  & ~~~Exponent & $d=2$ & $d=3$\\
\mr
$\nu_\parallel/\nu_\perp$ & 1.765(3) &  1.919(4) & ~~~$\delta$&0.452(1) & 0.756(1) \\
$\beta/\nu_\perp$& 0.799(2)& 1.394(5) & ~~~$\eta$ & 0.229(3) & 0.110(1) \\
$\gamma/\nu_\perp$& 0.401(4) & 0.212(96) & ~~~$z$ & 1.133(2) & 1.042(2)  \\
\br
\end{tabular}
\end{indented}
\end{table}

\section{\label{result}Results for $d=3$}

\begin{figure}[hbt]
\begin{center}
\includegraphics[width=11.5cm,clip=true]{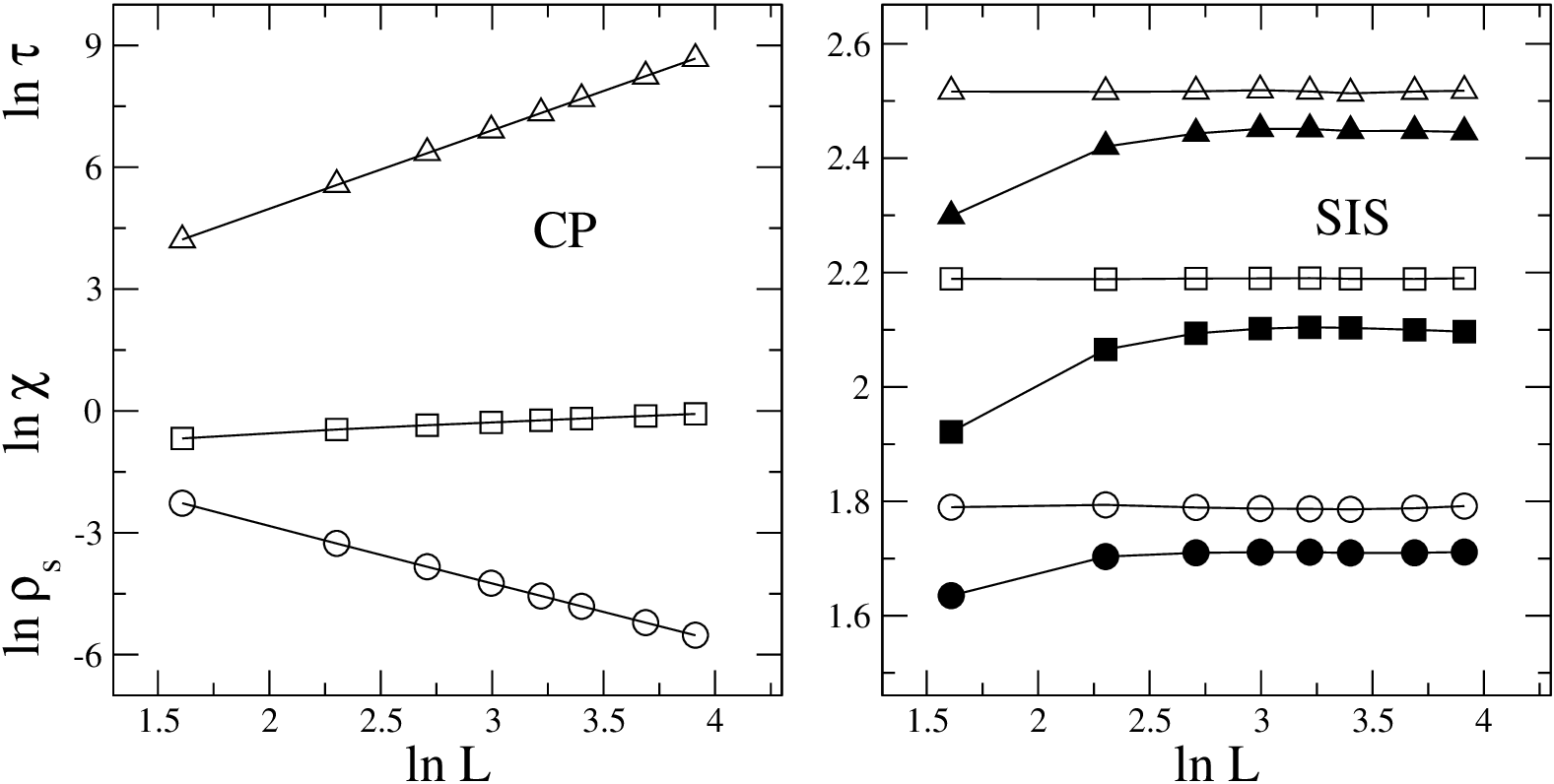}
\end{center}
\caption{\label{fig:qscp_a} Left: critical QS quantities against system size for three-dimensional CP. Lines are least square fits by (\ref{eq:correction}). Right: critical QS quantities for SIS rescaled by pure (filled symbols) and corrected (open symbols) power laws. Curves were shifted for sake of visibility.}
\end{figure}

Quasistationary simulations of the CP and SIS in $d=3$ dimensions are shown in figure \ref{fig:qscp_a}. The CP critical rate $\lambda_c = 1.31686(1)$ was taken from \cite{rewetting} whereas $\lambda_c = 0.24805(2)$ for the SIS model was estimated in the present work using spreading analysis. Least square fits for $L\ge20$ provide the exponents $\beta/\nu_\perp = 1.400(8)$, $\gamma/\nu_\perp = 0.233(5)$, and $\nu_\parallel/\nu_\perp = 1.919(9)$ for CP while $\beta/\nu_\perp = 1.395(5)$, $\gamma/\nu_\perp = 0.252(14)$, and $\nu_\parallel/\nu_\perp = 1.944(6)$ were obtained for SIS model. Notice that all of them are consistent with DP class, although data are deviated from pure power laws for small sizes. This effect can be taken in to account with a suitable correction to the amplitude of the scale law for lifetime given by \cite{Oliveira2005}
\begin{equation}
\label{eq:correction}
\ln \tau = \frac{\nu_\parallel}{\nu_\perp}\ln L + \frac{A}{L^{d\vartheta}} + \mbox{const.},
\end{equation}
and so on for the others QS quantities. {Actually, our results are not significantly affected by the particular choice of the correction}. The value $\vartheta = 0.75$ was established for CP in $d=1$ \cite{Oliveira2005} and also adopted  here for $d=3$. Using this correction, the exponents are $\beta/\nu_\perp = 1.395(4)$, $\gamma/\nu_\perp = 0.216(3)$, and $\nu_\parallel/\nu_\perp = 1.916(5)$ for contact process representing a significant reduction of the error estimates and an improvement of the closeness with DP class.  For the SIS model, the correction to  the scaling provides $\beta/\nu_\perp = 1.403(4)$, $\gamma/\nu_\perp = 0.209(2)$, and $\nu_\parallel/\nu_\perp = 1.922(2)$, an even better improvement of the estimates as well as their proximities with DP class. In left panel of figure \ref{fig:qscp_a}, the rescalings by pure power laws, $\tau / L^ {\nu_\parallel/\nu_\perp}$ and so on,  are compared with those given by (\ref{eq:correction}). It is neat the {overmatching} fit between data and the ansatz (\ref{eq:correction}).

\begin{figure}[hbt]
\begin{center}
\includegraphics[width=9.0cm,clip=true]{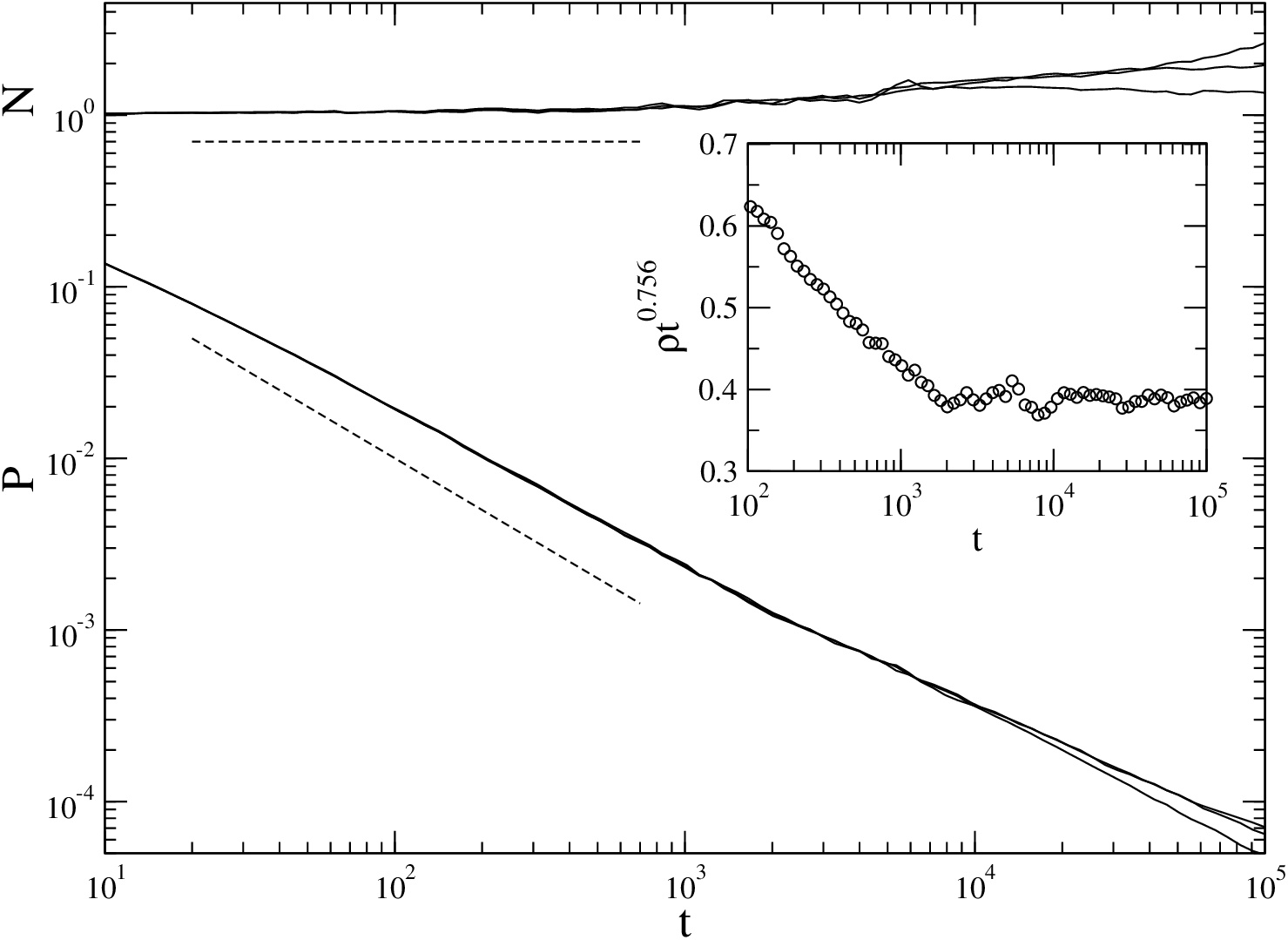}
\end{center}
\caption{\label{fig:fig2} Spreading analysis for the three-dimensional CRP around the critical point. Bottom and top groups of curves correspond to survival probability $P$ and mean number of occupied sites $N$. In each group, are shown curves for $\lambda = 1.00362,~1.00363,$ and 1.00364 from bottom to top. Dashed lines represent slopes 0 and -1  corresponding to the DP mean field exponents $\eta$ and $\delta$, respectively. Inset: mean critical density (not the QS one) rescaled by the power law $t^{-\delta}$ for a initial condition with all sites occupied.}
\end{figure}

A central point of the present work is the QS analysis of CRP in three-dimensions for which the universality class was left as an open question. Firstly, we recalculated the critical point determination using spreading analysis and found a value approximately 0.3\% larger than the originally reported \cite{Ferreira2005}. Using the criterion of downward and upward curvatures of the plots $N~vs.~t$, our best estimate is $\lambda_c=1.00363(1)$. This small discrepancy is central since this certainly excludes the equality between creation and annihilation critical rates. The curves $N(t)$ and $P(t)$ have transients consistent with the DP mean-field exponents (figure \ref{fig:fig2}), in agreement with the previously reported CRP simulations \cite{Ferreira2005}. Even though a large power interval was not obtained for the largest time simulated, the exponents fitted in the interval $t=10^4-10^ 5$, namely, $\eta = 0.09(2)$, $\delta = 0.78(3)$ and $z = 1.04(1)$, do not exclude the DP class (table \ref{exponents}). Additionally, the critical density for a fully occupied initial configuration decays as $\rho\sim t^{-0.76(1)}$  for $t>10^3$, in excellent agreement with DP class (inset of figure  \ref{fig:fig2}).

\begin{figure}[hbt]
\begin{center}
\includegraphics[width=10cm,height=!,clip=true]{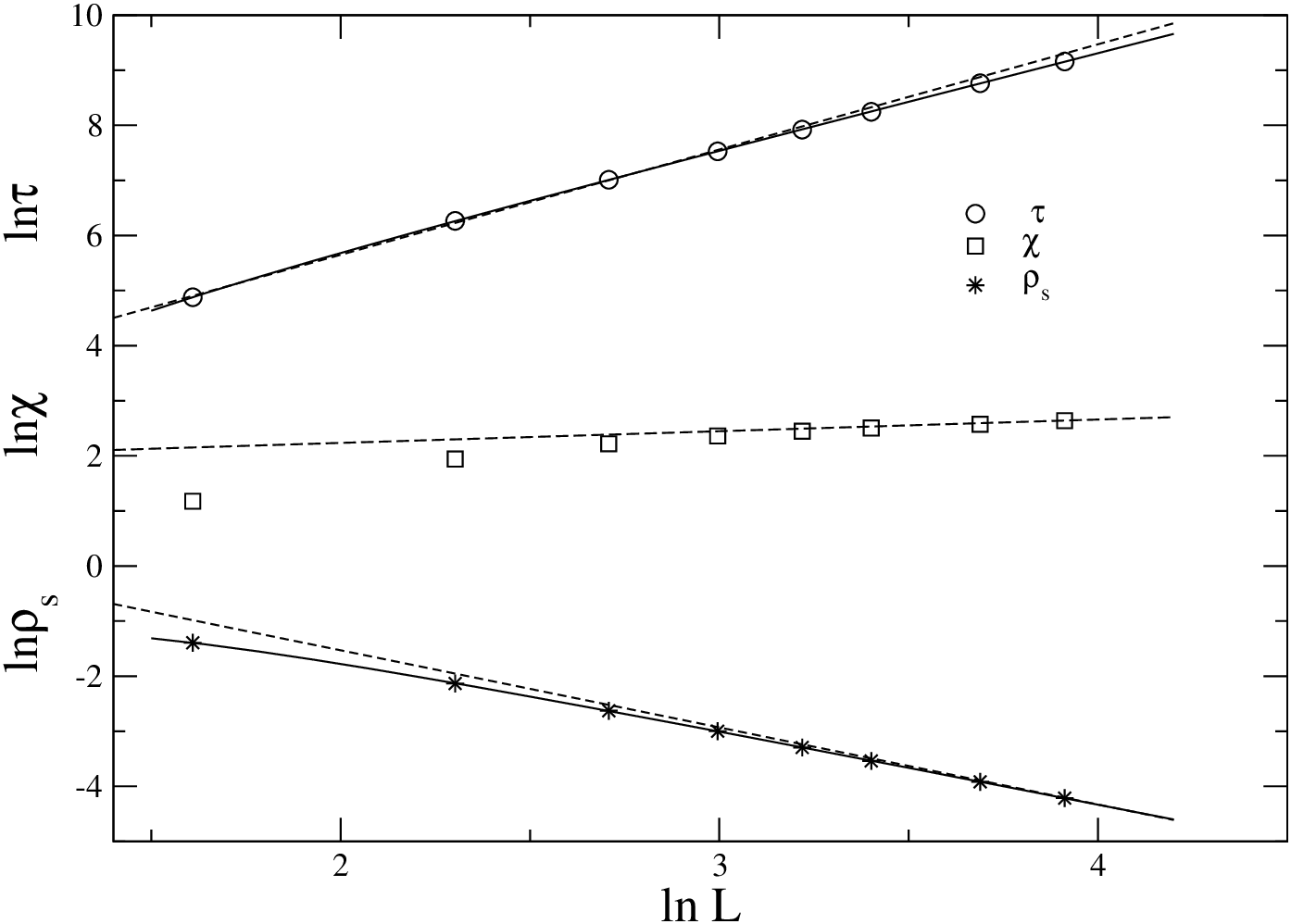} \\
\caption{\label{fig:fig3} QS simulations of CRP model at criticality. Solid lines are least square fits using (\ref{eq:correction}) and dashed lines are power laws with the DP exponents taken from table \ref{exponents}.}
\end{center}
\end{figure}

\begin{figure}
\begin{center}
\includegraphics[width=9cm,clip=true]{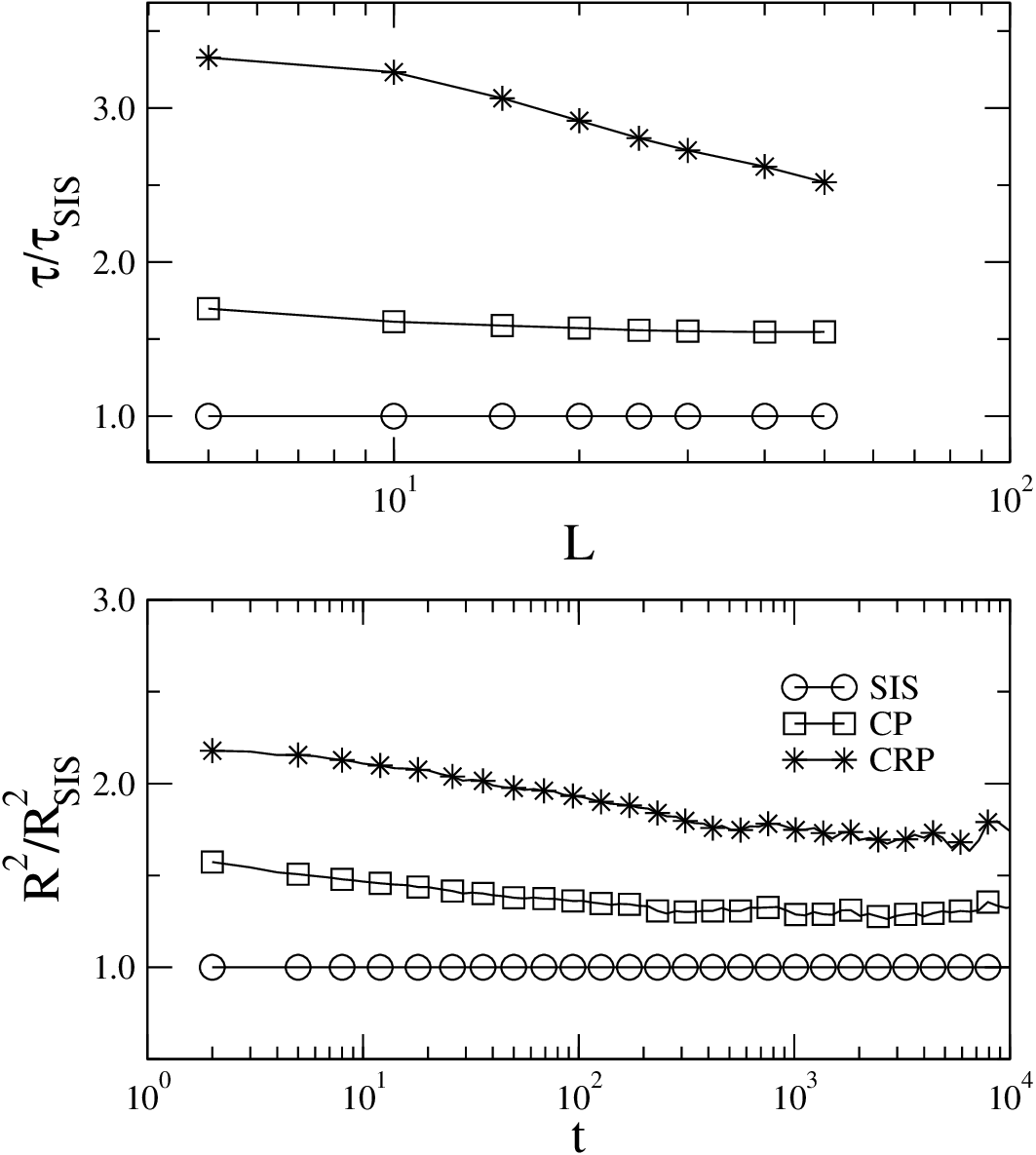}
\caption{\label{fig:A_CP_CRP} Critical quantities normalized by the SIS model. Top: lifetimes as functions of the system sizes. Bottom: squared spreading radius versus time. As one can see, diffusion and lifetimes are enhanced in CRP model.}
\end{center}
\end{figure}

Results of QS simulations for the CRP model at the critical point
are shown in Fig. \ref{fig:fig3}. The QS exponents obtained from
fits in the range $L>20$ are $\beta/\nu_\perp = 1.325(6)$,
$\gamma/\nu_\perp = 0.29(5)$, and $\nu_\parallel/\nu_\perp =
1.79(1)$ which are not concluding about the agreement with the DP
universality class. However, with exception of the lifetime, the
exponents seem converging to the DP value. Applying a correction to
the scaling (\ref{eq:correction}), the corrected exponents for
$\rho_s$ and $\chi$ , $\beta/\nu_\perp = 1.374(3)$ and
$\gamma/\nu_\perp = 0.19(4)$, are closer to DP class. The lifetime
analysis deserves some comments. The QS method used in present work,
involves the transition to the absorbing configuration passing by
the state with a single occupied site. In the critical CRP, the
creation event occurs with a frequency slightly larger than the
annihilation, which can be easily verified from the model critical
rate together with its rules. Consequently, large clusters of
occupied sites are much more seldom in CRP than in CP and, mainly,
in SIS models implying in a too small frequency of visiting the
pre-absorbing state (figure \ref{fig:A_CP_CRP}). Moreover, the
average time that a sample is kept in the list containing the system
history, given by $M/p_{rep}$, must be much larger than the sample
lifetime, $\tau=1/p_1$, to avoid a same run anomalously contributing
many times to the list of configurations \cite{Oliveira2005}.
Obviously, this effect is enhanced as the system size increases.
Thus, a computationally prohibitive small $p_{rep}$ and/or large $M$
together with too large relaxation and averaging times are demanded.
Instead of lifetime, we can determine the characteristic time
$\tau_\rho$ to the density reaches the QS state in conventional QS
simulations \cite{Marro,Ferreira2004}. The scale relations $\rho\sim
L^{1.41(2)}$ and $\tau_\rho \sim L^{1.88(2)}$ were found for CRP, in
agreement with the DP class as predicted by Janssen-Grassberger
conjecture.

\begin{figure}
\includegraphics[width=13cm,height=!,clip=true]{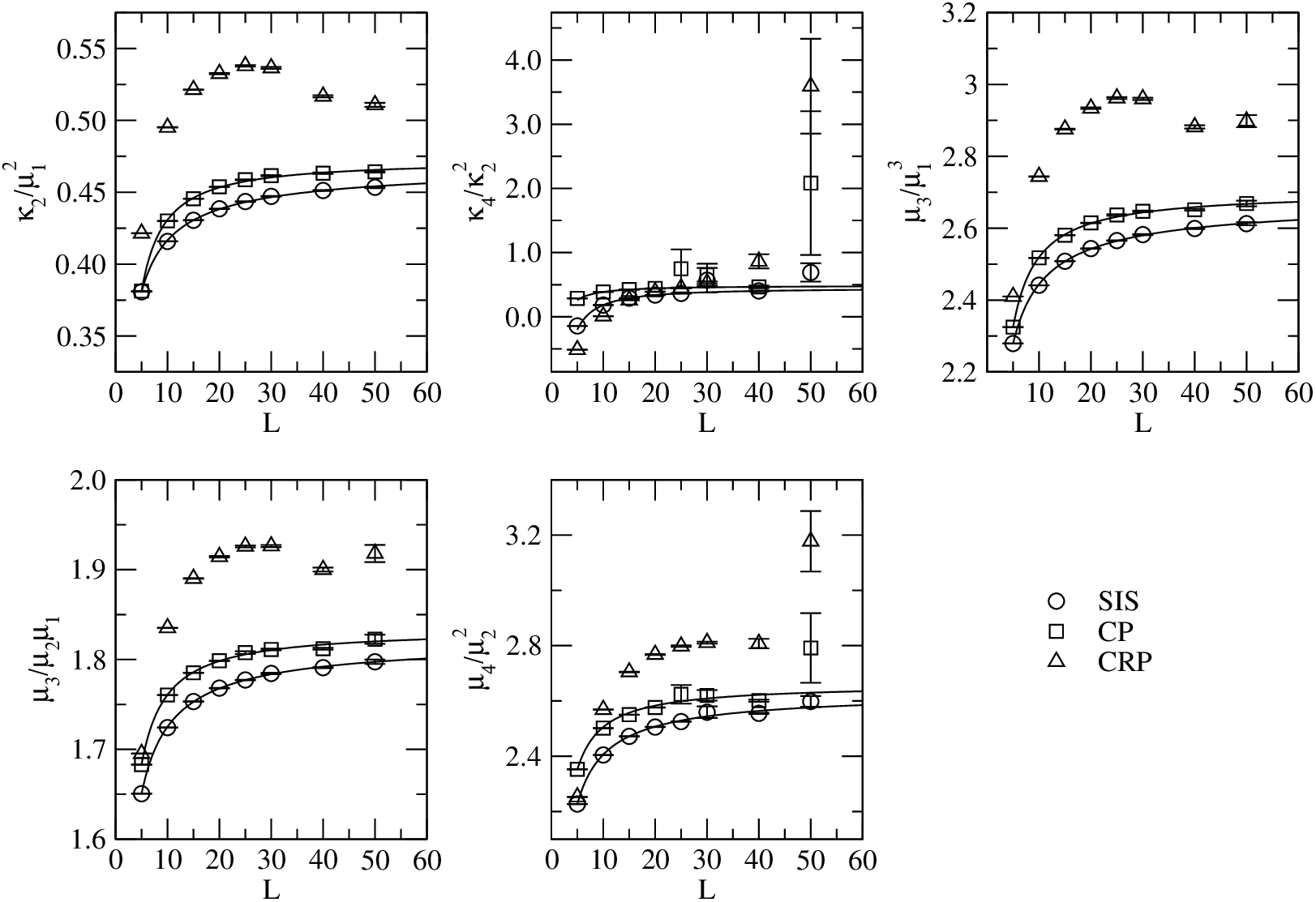}
\caption{\label{fig:moments} Moment rations for SIS, CP and CRP models in $d=3$ dimensions. Solid lines are non-linear fits by function $m(L) = m(\infty)+aL^{-\psi}$ with a correlation coefficient $r>0.999$.}
\end{figure}

{Moments and cumulants for DP class in $d=1-5$ with a homogeneous constant external source were recently reported by Janssen \textit{et al}. \cite{Janssen2007}.} So, let us introduce the notation $\mu_n = \langle\rho^n\rangle$ and $\kappa_n$ to the $n$th moment and cumulant of the order parameter, respectively. In particular, we have
\begin{equation}
\kappa_2 =  \mu_2-\mu_1^2
\end{equation}
and
\begin{equation}
\kappa_4 = \mu_4-4\mu_3\mu_1-3\mu_2^2+12\mu_2\mu_1^2-6\mu_1^4.
\end{equation}
Dickman and da Silva \cite{jaff} showed that the ratios $\kappa_2/\mu_1^2$, $\kappa_4/\kappa_2^2$, $\mu_3/\mu_1^3$, $\mu_3/\mu_1\mu_2$, and $\mu_4/\mu_2^2$ assume universal values for DP class in $d=1$ and $d=2$ dimensions. It is important to notice that both even and odd moments can be studied, since the order parameter is non-negative. Several moment and/or cumulant ratios as functions of the system size for $3d$ models are shown in figure \ref{fig:moments}. The moment ratios for SIS and CP models exhibit finite size effects, but seem to monotonically converge to a constant value when $L\rightarrow\infty$. Assuming a correction given by $m(L) = m(\infty)+aL^{-\psi}$, where $\psi$ and $a$ are fit parameters, the asymptotic moment ratios can be estimated. Since some data for fourth moment were ruled by very large error bars, these data were excluded from the fits. Extrapolated moment ratios for CP and SIS are listed in table \ref{moments} and are consistent with the hypothesis of universality. {It is worth to stress that ratios shown in table \ref{moments} differ from those reported by Janssen \textit{et al}. \cite{Janssen2007} since, in our studies, there is no external source. The same occurs in $d=1$ and $2$ when the results of Janssen \textit{et al}. \cite{Janssen2007} are compared with those from reference \cite{jaff}.}

CRP seems to be different from CP and SIS. Indeed, instead of the monotonic convergence to a constant value, the moment ratios decreases after a maximum. This difference can be again associated to the proximity between critical creation and annihilation events and the resultant high diffusivity. Actually, the ratio $\mu_2/\mu_1^2 = \kappa_2/\mu_1^ 2+1$ first grows towards the value 1.550. This value is nearby the established value of 1.660 for CP on a complete graph \cite{Vidigal}. Again, we have a mean-field behaviour for small system that must converge to the usual DP for asymptotic large systems.
\begin{table}[hbt]
\caption{\label{moments} Asymptotic moment ratios for SIS and CP in $d=3$.}
\begin{indented}
\item[]\begin{tabular}{cccccc}
\br
Model& $\kappa_2/m_1^2$ &  $\kappa_4/\kappa_2^2$  &  $\mu_3/\mu_1^3$ & $\mu_3/\mu_1\mu_2$ & $\mu_4/\mu_2^2$\\
\mr
SIS & 0.469(3) &  0.490(6)  & 2.678(12) & 1.822(3) & 2.629(12) \\
CP  & 0.470(2) &  0.454(10) & 2.697(6)  & 1.833(3) & 2.649(18) \\
\br
\end{tabular}
\end{indented}
\end{table}

\section{\label{conclusion}Conclusions}

We performed large-scale simulations of the contact process, the
susceptible-infected-susceptible and the contact replication process
in three dimensions. Applying the quasi-stationary simulation
method, we were able to determine the moment ratios of the order
parameters for DP class in three dimensions {in the absence of an external field}. We also show that the
mean-field exponents in $d=3$ for the CRP, reported in
\cite{Ferreira2005} are a transient observed in the spreading
analysis.

Quasi-stationary simulations and suitable corrections to the scaling
revealed that the CRP model belongs to the directed percolation
universality class, as expected by the Janssen-Grassberger criterion.
However, the moment ratios for CRP agree with the universal DP
values in $d=1$  and 2 dimensions but do not in $d=3$. The
discrepancy lies on the closeness between critical creation and
annihilation  events, which gives rise to a diffusive transient
behaviour in the CRP  model, also responsible by the transient mean
field exponents obtained in $d=3$.

\ack

This work was supported by the Brazilian agencies FAPEMIG and CNPq.

\end{document}